\input epsf
\def\square{\hbox{{$\sqcup$}\llap{$\sqcap$}}}   

\documentstyle[aps]{revtex}
\begin{document}


\title{Colliding Axion-Dilaton Plane Waves from Black Holes}
\author{Patricia Schwarz
\footnote{
http://www.theory.caltech.edu/people/patricia}
}
\address{Caltech 452-48}
\address{Pasadena CA 91125}
\date{July, 1997}
\maketitle

\begin{abstract}
The colliding plane wave metric discovered by Ferrari and Iba\~{n}ez 
to be locally isometric to the interior 
of a Schwarzschild black hole is extended
to the case of general axion-dilaton black holes.
Because the transformation maps either black hole horizon to
the focal plane of the colliding waves, this entire class
of colliding plane wave spacetimes only suffers from the
formation of spacetime singularities in the limits where
the inner horizon itself is singular, which occur in the
Schwarzschild and dilaton black hole limits. 
The supersymmetric limit corresponding to the extreme axion-dilaton
black hole yields the Bertotti-Robinson metric with the axion and
dilaton fields flowing to fixed constant values. 
The maximal analytic extension of this metric across
the Cauchy horizon yields a spacetime in which
two sandwich waves in a cylindrical universe collide to produce a
semi-infinite chain of Reissner-Nordstrom-like wormholes.
The focussing of particle and string geodesics in this
spacetime is explored. 
\end{abstract}

\section{Introduction}
A static, spherically symmetric black hole spacetime is only
static in the regions where ${\partial \over \partial t}$ is
timelike. In the trapped region between the two
horizons, the metric is quite violently
dependent on the timelike radial coordinate, while 
${\partial \over \partial t}$ and 
${\partial \over \partial \phi}$ act as a pair of
purely spacelike commuting Killing vectors. A violently
time-dependent spacetime with two commuting
spacelike Killing vectors is also a potential description
of the spacetime of two colliding plane symmetric
gravitational waves. This idea was first recognized and 
explored in the in the Einstein-Maxwell limit by
Chandrasekhar in 1984 \cite{Chan}. 
A colliding plane wave metric locally isometric to the 
interior  of a Schwarzschild 
black hole was obtained by Ferrari, Iba\~{n}ez and Bruni 
in 1987 \cite{FerIban1,FerIban2}. The direct transformation from
a Schwarzschild black hole to a colliding plane wave spacetime
was described by Yurtsever in 1988 \cite{Yurt4,Yurt1}.

The purpose of this paper is to extend this analysis to the case
of axion-dilaton black holes \cite{RK1,RK2,RK3,RK4}
that are N=4 supersymmetric solutions of low energy string
theory, and to compare string and particle propagation in
the resulting spacetimes.  

In section II, we display the transformation
between a Schwarzschild trapped region and a colliding 
plane wave spacetime elucidated by Yurtsever. Then
we extend this transformation to the general case
of axion-dilaton black holes found in low-energy string theory.
We show how these plane
wave collisions end in the formation of singularities $only$
when they represent transformations of black hole spacetimes 
where the singularity is touching the trapped region, as in 
the case of the Schwarzschild and the singular dilaton black 
holes.

The non-singular colliding wave spacetimes have Killing-Cauchy
horizons instead of singularities. The curvature at the
Killing-Cauchy horizon is equal to the curvature of the appropriate 
black hole horizon locally isometric to that particular plane 
wave spacetime. The metric can be extended across this horizon
in an intuitively appealing manner, but the price of avoiding
the singularity is the loss of global hyperbolicity. This is
consistent with Hawking-Penrose singularity theorems in that
geodesic focusing forces a choice between a local or 
global pathology. 

In section III we review the
work done by Yurtsever on the asymptotic
structure of colliding plane wave spacetimes, and we
show where the transformed black hole solutions fit in
this general classification scheme.

In section IV we compare the asymptotic causal structure of
axion-dilaton colliding plane waves with that of general
colliding plane wave solutions of the vacuum Einstein and
Maxwell-Einstein-dilaton equations.

In section V we show that the maximal analytic extension of the
general axion-dilaton colliding plane wave spacetime is two
collinearly polarized waves propagating in a cylindrical universe
to create a black hole with the same causal structure as
an infinite chain of wormholes in Reissner-Nordstrom spacetime.
In the event that either of the incoming waves has a delta function profile 
in the incoming regions, the maximal analytic extension degenerates
to the extreme dilaton supersymmetric configuration with $1/2$
of $N=4$ supersymmetry unbroken.

In section VI we compare particle and string 
propagation in an exact plane
wave background and plot the effects of violation of the Principle
of Equivalence by strings. We also briefly examine the
issue of more realistic finite-sized almost-plane waves.

\section{Colliding Waves Out of Black Holes}

\vskip .25 cm
\centerline{\epsfxsize=3.0in\epsfbox{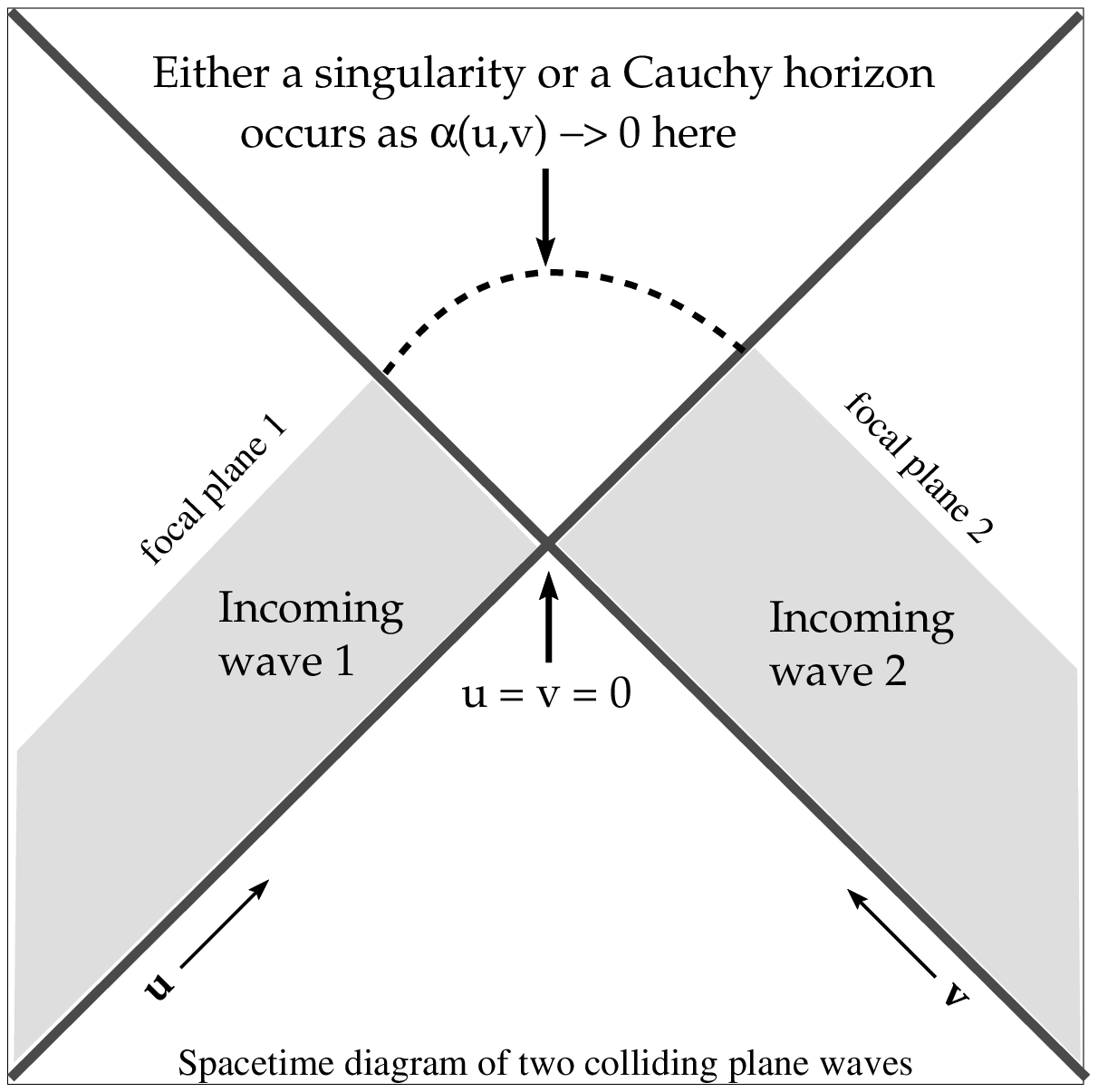}}
{\footnotesize {\sl Fig.1} Spacetime diagram of two
colliding plane waves}
\vskip 0.25 cm

Inside the trapped region ($r\le 2 M$)
of a Schwarzschild black hole
the metric can be written:

\begin{equation}\label{hole}
ds^2 = - {r \over (2 M - r)}\,dr^2 + r^2\,d\theta^2 + 
{(2 M - r) \over r}\,dt^2 + r^2 \sin^2 \theta\,d\phi^2.
\end{equation}

\noindent On the other hand, the
metric for the interaction region of two colliding,
collinearly polarized, plane symmetric gravity waves can be
written in the form:

\begin{equation}\label{form}
ds^2 = - e^{-M(u,v)}\,du\,dv  + e^{-U(u,v)} \left(
{e^{V(u,v)}\,dx^2 + e^{-V(u,v)}\,dy^2} \right).
\end{equation}

The former spacetime can be put in the form of the
latter using the coordinate transformation \cite{Yurt4,Yurt1}

\begin{equation}
r \to M(1 - \sin(u + v)),\quad\theta \to {\pi\over 2} 
+ v - u,\quad t \to x,\quad \phi \to 1 + {y\over M},
\end{equation}
 
\noindent for $u\ge 0, v\ge 0,$ and $u + v\le \pi/2.$
To make a colliding plane wave spacetime we have to 
analytically continue $y$ past the
cyclic boundary conditions on $\phi.$ Therefore the
resulting metric

\begin{eqnarray}\label{yurt1}
ds^2 = - 4 M^2\, {{\left( 1 - \sin (u + v) \right)}^2}\,du\,dv 
&+& {{\cos^2 (u + v)}\over {(1 - \sin (u + v)})^2}\,dx^2\nonumber\\
&+& {{{\cos^2 (u - v)}}\,{\left( 1 - \sin (u + v) \right)}^2}\,dy^2
\end{eqnarray}

\noindent is locally, but not globally, isometric to
(\ref{hole}).

There is one slight problem with this metric: it serves as
a good description of the interaction region 
for two colliding
plane waves, but it doesn't describe the spacetime before
the two waves have met. Penrose and Khan \cite{KhanPen} came up with an
effective yet slightly flawed prescription for constructing
incoming waves from a metric for a colliding wave
interaction region: replace $u$ and $v$, respectively, by
$u H(u)$ and $v H(v),$ where $H(x)$ is the Heaviside step 
function. Thus an incoming wave in the region $u>0,v<0$ can
be written

\begin{equation}\label{minsw}
ds^2 = - 4 M^2\, {{\left( 1 - \sin (u) \right)}^2}\,du\,dv 
+ {{\cos^2 (u)}\over {(1 - \sin (u)})^2}\,dx^2
+ {{{\cos^2 (u)}}\,{\left( 1 - \sin (u) \right)}^2}\,dy^2.
\end{equation}

Length scales  for the colliding wave system are introduced via 
$u \to u/a, v \to v/b.$ The requirement that the metric be continuous
with flat spacetime at $u=v=0$ relates the
focal lengths and amplitudes of the incoming waves 
through $a b = 4 M^2,$  which we will see more of later.
The spacelike Killing vector
${\partial \over \partial x}$ becomes singular when ${u\over a}
+ {v \over b} \to {\pi\over 2},$ and moreover, 
$R_{\mu\alpha\nu\beta} R^{\mu\alpha\nu\beta} \to \infty$
there, indicating a spacetime singularity.
The incoming wave metrics
obtained by the above Khan-Penrose prescription have
$R_{\mu\alpha\nu\beta} R^{\mu\alpha\nu\beta}= 0$ but the
Weyl tensor component 

\begin{equation}\label{cuxux}
C_{uxux} = 
{{-3\,{{\left( \cos ({u\over {2\,a}}) + 
          \sin ({u\over {2\,a}}) \right) }^2}}\over 
   {{a^2}\,{{\left( \cos ({u\over {2\,a}}) - 
          \sin ({u\over {2\,a}}) \right) }^4}}}
\end{equation}
 
\noindent blows up as $u \to \pi a/2,$ showing that the incoming
waves are singular in some sense before they collide.

A better-behaved metric  
can be obtained by sending $(u,v)$ to $(-u,-v):$ 

\begin{eqnarray}\label{yurt2}
ds^2 = - 4 M^2\, {{\left( 1 + \sin (u + v) \right)}^2}\,du\,dv 
&+& {{\cos^2 (u + v)}\over {(1 + \sin (u + v)})^2}\,dx^2\nonumber\\
&+& {{{\cos^2 (u - v)}}\,{\left( 1 + \sin (u + v) \right)}^2}\,dy^2.
\end{eqnarray}

This metric is also
locally isometric to the trapped region of a Schwarzschild
black hole, except that 
$R_{\mu\alpha\nu\beta} R^{\mu\alpha\nu\beta} = 3/{4 M^4}$ 
in the limit
${u\over a} + {v \over b} \to {\pi\over 2}$. The spacelike
Killing vector ${\partial \over \partial x}$ becomes
null there, signalling a Cauchy horizon (because initial
data that is spatially homogenous in the $x$-direction
ceases to be so when ${\partial \over \partial x}$ is
no longer spacelike).
The incoming waves extended from this collision region have

\begin{equation}
C_{uxux} = 
{{-3\,{{\left( \cos ({u\over {2\,a}}) - 
          \sin ({u\over {2\,a}}) \right) }^2}}\over 
   {{a^2}\,{{\left( \cos ({u\over {2\,a}}) + 
          \sin ({u\over {2\,a}}) \right) }^4}}},
\end{equation}

\noindent which vanishes on the incoming focal plane
$u = {\pi a \over 2}.$
These waves are called ``sandwich waves," the curvature
being neatly sandwiched between the past wave front and the focal plane
to the future. The incoming waves in (\ref{minsw}) are not sandwich
waves in this sense.

\subsection{Axion-Dilaton Black Holes}

In order to better understand this pattern of singular
and nonsingular behavior, we will extend the coordinate transformation made for the Schwarzschild black hole to the general case of an axion-dilaton 
black hole in $d=4$ with $N$ U(1) gauge fields, with the action

\begin{equation}\label{action}
S_{eff} = {1\over 16\pi}\,\int d^4 x
\, \sqrt{-g}\,\, (
- R  + {1\over 2} {\partial_\mu \lambda \partial^\mu \bar{\lambda} \over 
({\rm Im}\lambda)^2} 
- \sum_{n=1}^{N}{F^{(n)}_{\mu\nu} {}^\star \tilde{F}^{(n)}{}^{\mu\nu}}),
\end{equation}

\noindent where 
$ \tilde{F}^{\mu\nu} = e^{-2 \phi} {}^\star F^{\mu\nu} - i \psi F^{\mu\nu}$. The axion $(\psi)$ and dilaton $(\phi)$
fields are combined into
$\lambda = \psi + i e^{-2 \phi},$ and ${}^\star$ is the spacetime
dual operation. 

In the trapped region the metric can be written

\begin{eqnarray}\label{axibh}
ds^2 = - {{(r^2 - |\Upsilon|^2)}\over (r_+ - r)(r - r_-)}\,dr^2 
+ (r^2 - |\Upsilon|^2)\,d\theta^2 &+& 
{(r_+ - r)(r - r_-)\over (r^2 - |\Upsilon|^2)}\,dt^2\nonumber\\
&+& (r^2 - |\Upsilon|^2) \sin^2{\theta}\,d\phi^2,
\end{eqnarray}

\noindent where

\begin{equation}\label{rodef}
r_\pm = M \pm r_0, \quad r_0^2 = M^2 + |\Upsilon|^2 - 
4 \sum{|\Gamma^{(n)}|^2},\quad \Gamma^{(n)} = {1 \over 2}
(Q^{(n)} + i P^{(n)}),
\end{equation}

\noindent and

\begin{equation}\label{sigdelt}
\Upsilon = \Sigma - i \Delta 
= - {2 \over M} \sum_{n=1}^{N}{(\Gamma^{(n)})^2}.
\end{equation}

\noindent $\{Q^{(n)},P^{(n)}\}$ are the U(1) electric and magnetic 
charges, respectively. The entropy of the axion-dilaton black
hole is given by $1/4$ of the area of the horizon

\begin{equation}
S = {A \over 4} = \pi ((r_+)^2 - |\Upsilon|^2).
\end{equation}

It is important to remember that the coordinate $r$ is now measuring
time, so this is a highly time-dependent spacetime, not the
placid exterior of a classical black hole. In the extreme limit of
$r_+ \to r_- \to M,$ or $r_0 \to 0,$ the region over which $r$
is timelike shrinks to zero, and so the amount of violent time-dependence
inside the black hole shrinks away as well. The area of the
extreme black hole is

\begin{equation}
S_{extr} = {A_{extr} \over 4} = \pi (M^2 - |\Upsilon|^2).
\end{equation}

In general these axion-dilaton black holes have fascinating properties
and relationships to deep symmetries in string theory. \cite{RK1,RK2,RK3,RK4}
The parameter $r_0$ measures how far the black hole is 
from the extremal limit 
$r_+ = r_-,$ where the trapped region threatens to vanish and
reveal a naked singularity to the universe. The parameter $r_0$ also
measures the breaking of supersymmetries in the N=4 supergravity theory
underlying the action (\ref{action}). The condition $r_0 = 0$
corresponds to the saturation of the SUSY bound via 
$M = |z_1| > |z_2|$ or $M = |z_2| > |z_1|$
between the black hole mass and the largest of the eigenvalues
$(z_1,z_2)$ of the central charge matrix of the $N=4$ theory,
restoring $1/4$ of the broken $N=4$ supersymmetry. 
The area of the extreme horizon is proportional to the square
of the largest central charge at the ``fixed point" where the
other central charge vanishes.

The full saturation 
$M = |z_1| = |z_2|$ restores $1/2$ of the broken $N=4$
supersymmetry. The $r_0 \to 0$ limit of the
corresponding black hole is an extreme dilaton black hole 
with $M = |\Upsilon|,$ zero entropy and a singular horizon.
Hence supersymmetry serves as a cosmic censor for these black
holes as long as not more than $1/4$ of the $N=4$ supersymmetry
is restored.

The axion and dilaton fields add to this interesting behavior
at the horizon in the $r_0 = 0$ limit. 
At the extreme horizon they lose all dependence on their values
$\lambda_0 = \psi_0 + i e^{- 2 \phi_0}$
at spatial infinity and depend only on the values of quantized
conserved charges. 
For a single extreme black hole of
this type with $N$ electric and magnetic charges 
$Q^{(i)} + i P^{(i)} = e^{\phi_0}\,( n_i - \bar{\lambda_0} m_i),$
with $(n_i,m_i) \in Z,$ the axion and dilaton fields 
at the horizon reduce to \cite{RK5}

\begin{equation}\label{fixed}
\psi_f = {\sum{n_i m_i} \over \sum{m_i^2}},\quad
e^{- 2 \phi_f} = {(\sum\limits_{i<j}{(n_i m_j - n_j m_i)^2)^{1/2}}
 \over \sum{m_i^2}}.
\end{equation}

\vskip 0.25 cm
\centerline{\epsfxsize=3.0in\epsfbox{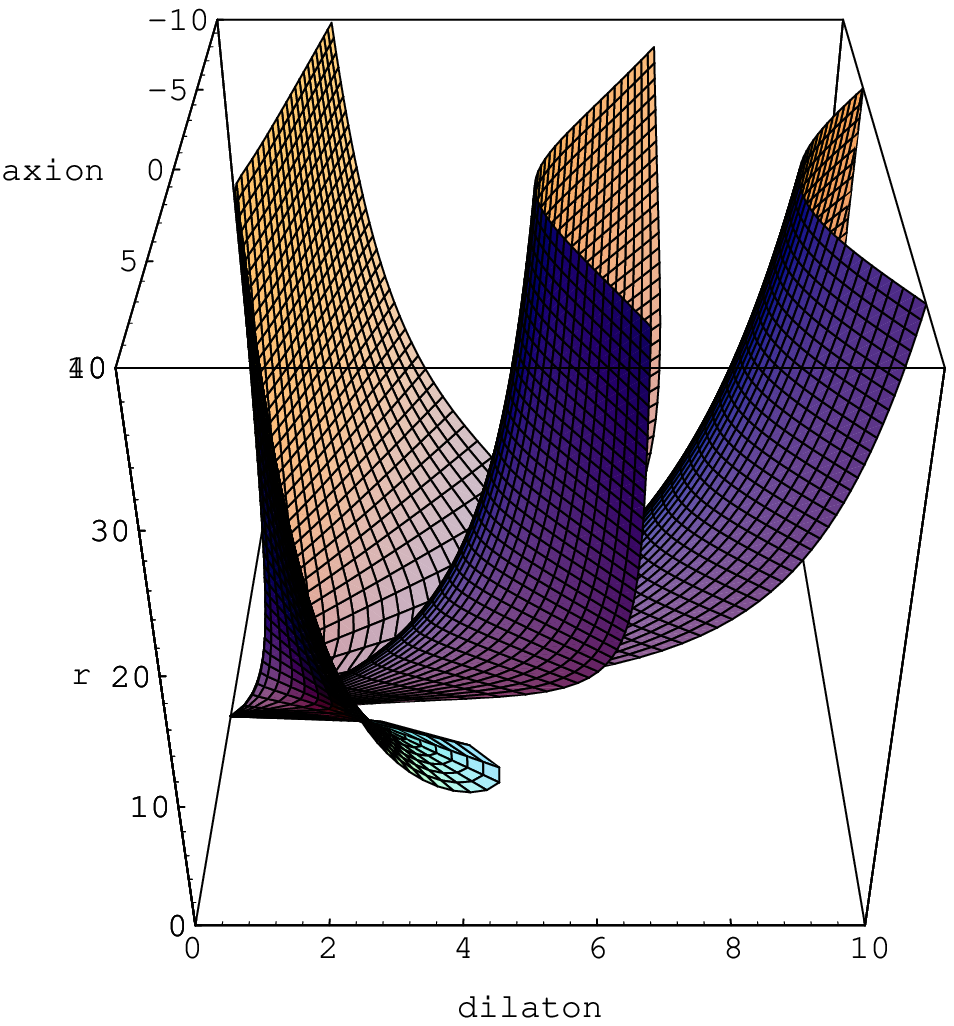}}
\noindent {\footnotesize {\sl Fig.2} These contours of
constant $\phi_0$ show how the axion and dilaton fields lose their
dependence on $\phi_0$ and $\psi_0$ and flow to fixed values on the
horizon of an extreme axion-dilaton black hole. Here the coordinate $r$
measures distance from the extreme horizon.}
\vskip 0.25 cm

\subsection{Axion-Dilaton Colliding Waves}

The coordinate transformation from $(r,\theta,t,\phi)$ to 
$(u,v,x,y)$ gives $r(u,v) \to r_{\pm}$ for the $(\pm)$
branch of the solution as $u/a + v/b \to \pi/2:$

\begin{eqnarray}\label{ctrans}
r &\to& M \pm r_0\, \sin ({u\over a} + {v\over b}),\qquad
\theta \to {\pi \over 2} \pm \left(
{u \over a} - {v \over b}\right), \label{trans}\\
t &\to& x\,{r_0}/{({{M^2}- {|\Upsilon|^2}} )^{1 \over 2}},
\qquad
\phi \to 1 +  y/{({{M^2}- {|\Upsilon|^2}} )^{1 \over 2}}.
\nonumber
\end{eqnarray}

\noindent The trapped region of the black hole is only locally 
isometric to the interaction region of the colliding plane wave 
spacetime, because we're sending the cyclic coordinate $\phi$
to the non-cyclic coordinate $y,$ to represent a plane wave infinite
in both the $x$ and $y$ directions.

The axion-dilaton black hole metric in the trapped region 
now takes the form 

\begin{eqnarray}\label{adcwmetric}
g_{uv} &=& {{-2\,\left({{{\left( M \pm 
          r_0\,\sin ({u\over a} + {v\over b})
           \right) }^2} -{|\Upsilon|^2}} \right)}
\over a b},\\
g_{xx} &=& {{\left({{ M^2} - {|\Upsilon|^2}}\right)\,
{{\cos ({u\over a} + {v\over b})}^
       2}}\over 
   {{{\left( M \pm 
           r_0\,\sin ({u\over a} + {v\over b})
           \right) }^2} -{|\Upsilon|^2} }}\nonumber\\
g_{yy} &=& {\cos^2 ({u\over a} - {v\over b})}\,
     {\left( {{{\left( M \pm 
          r_0\,\sin ({u\over a} + {v\over b})
           \right) }^2} -{|\Upsilon|^2} } \right) \over 
   {{ M^2} - {|\Upsilon|^2}}}.\nonumber
\end{eqnarray}

Requiring $g_{\mu\nu}=\eta_{\mu\nu}$ for $u=v=0$ 
constrains the incoming parameters $a$ and $b$ to satisy

\begin{equation}
a b = 4 (M^2 - |\Upsilon|^2) = {4 S_{extr} \over \pi}.
\end{equation}

\noindent Thus only when the mass is larger than the 
axion-dilaton charge and the entropy of the relevant black
hole is not vanishing are both parameters $a$ and $b$ 
non-vanishing. This constraint
is significant because the condition $M > |\Upsilon|$ is a supersymmetry
bound that helps enforce cosmic censorship in the black hole system.
This bound in the colliding plane wave system tells us that
the effective focal length of the colliding axion-dilaton 
plane wave system is not negative
($f = \pi \sqrt{a b}/2 \ge 0$) and only approaches zero in the
singular extreme dilaton limit $M = |\Upsilon|.$ This looks and acts like a 
supersymmetric enforcement of cosmic censorship, although 
in the context of the colliding-wave problem,
it was derived by 
requiring that the spacetime be exactly flat before the arrival of each 
incoming wave. 

Abbreviating $r_{\pm}(u,v) = M \pm r_0\, \sin(u/a + v/b),$ the axion and
dilaton fields become

\begin{equation}
\psi(u,v) = {\psi_0\, (\Delta^2 + (\Sigma + {r_{\pm}(u,v)})^2) 
- 2 e^{-2 \phi_0} \Delta\, r_{\pm}(u,v)
\over 
\Delta^2 + (\Sigma + r_{\pm}(u,v))^2}
\end{equation}

\noindent and

\begin{equation}
e^{-2 \phi(u,v)}\, = e^{- 2 \phi_0}\,\, {{r_{\pm}(u,v)}^2 - (\Sigma^2 + \Delta^2) 
\over \Delta^2 + (\Sigma + r_{\pm}(u,v))^2},
\end{equation}

\noindent where $\Sigma$ and $\Delta$ are given by
equation (\ref{sigdelt}).

Transforming from $(t,\phi)$ to $(x,y)$ by (\ref{ctrans}), the
$N$ U(1) potentials with electric and magnetic charges $(Q^{(n)}, P^{(n)})$
are transformed from 
$(A^{(n)}_t\,,A^{(n)}_\phi)$ to 

\begin{equation}
A^{(n)}_x = {{e^{{\phi_0}}\,r_0\,
      \left( {P^{(n)}}\,{\Delta} + {Q^{(n)}}\,(\Sigma + r_{\pm}(u,v))\right) }
     \over {{\sqrt{{M^2} - {|\Upsilon|^2}}}\,
      \left( {{r_{\pm}(u,v)}^
          2} -{{{\Delta}}^2} - {\Sigma^2} \right) }},
\end{equation}

\begin{equation}
A^{(n)}_y \,= - {{{e^{\phi_0}}\,P^{(n)}\,
      \cos ({u\over a} - {v\over b})}\over 
    {{\sqrt{{M^2} - {|\Upsilon|^2}}}}}.
\end{equation}

The value of $R_{\mu\nu\rho\lambda} R^{\mu\nu\rho\lambda}$
in the limit $u/a + v/b \to \pi/2$ is equal to value
of $R_{\mu\nu\rho\lambda} R^{\mu\nu\rho\lambda}$ for the
equivalent axion-dilaton black hole, evaluated at
$r = r_{\pm} = M \pm r_0.$ One can see from
the following equation that this 
quantity will only blow up in two limits: the Schwarzschild
limit $|\Upsilon| = 0,\,r = r_- = 0$ and the extreme dilaton
limit $r_- = |\Upsilon|:$

\begin{equation}
R_{\mu\nu\rho\lambda} R^{\mu\nu\rho\lambda} =
{{8\,\left( {M^4} + 4\,{M^2}\,{{r_0}^2} + 
       12\,M\,{{r_0}^3} + 7\,{{r_0}^4} - 
       2\,{M^2}\,{|\Upsilon|^2} + 
       2\,{{r_0}^2}\,{|\Upsilon|^2} + {|\Upsilon|^4} \right) }
    \over {(r_{\pm}^2 - |\Upsilon|^2)^4}}
\end{equation}

\subsection{Extreme Limit of Axion-Dilaton Colliding Waves}

For axion-dilaton black holes the limit 
$r_0 \to 0$ corresponds to
the apparent vanishing of the trapped region between 
$r_+ = M + r_0$ and
$r_- = M - r_0.$ 
This also corresponds to the restoration of $1/4$ of the
broken $N=4$ supersymmetry in the background 
supergravity theory
and fixed values for the axion and dilaton fields at the extreme
horizon. For axion-dilaton colliding waves the 
$r_0 \to 0$ limit
gives the Bertotti-Robinson colliding plane wave spacetime

\begin{equation}\label{bert}
ds^2 = - du\,dv + {{\cos^2({u\over a} + {v\over b})}}\,
dx^2 + {\cos^2 ({u\over a} - {v\over b})}\,dy^2.
\end{equation}

In this limit the axion and dilaton fields reduce to

\begin{equation}
\psi_f = {\psi_0\, (\Delta^2 + (\Sigma + M)^2) 
- 2 e^{-2 \phi_0} \Delta\, M
\over 
\Delta^2 + (\Sigma + M)^2},\quad
e^{-2 \phi_f}\, = e^{- 2 \phi_0}\,\, {M^2 - (\Sigma^2 + \Delta^2) 
\over \Delta^2 + (\Sigma + M)^2},
\end{equation}

\noindent and reduce to (\ref{fixed}) when
written in terms of the Dirac-quantized conserved charges. 
The axion and dilaton are constant and take their critical 
values over the entire
Bertotti-Robinson spacetime, even in the flat region before
either wave has passed. Note that the axion and dilaton fields 
for $r_0 \ne 0$ also take their fixed
constant values in the flat region before the waves have arrived, but
evolve to their values at $r_{\pm}$  on the focal planes of the
incoming and colliding waves.

The incoming wave obtained from the above Bertotti-Robinson metric 
via the Khan-Penrose prescription for $u > 0, v < 0$
is
\begin{equation}\label{inwave}
ds^2 = - du\,dv + \cos^2({u\over a})\,
dx^2 + \cos^2({u\over a})\,dy^2.
\end{equation}

\noindent (The other incoming wave is the same as
above with 
$u \to v, a \to b$.) Using the coordinate transformation

\begin{eqnarray}
u &=& U\\
v &=& V - {1 \over a} \tan({u \over a}) (X^2 + Y^2)\nonumber\\
x &=& X/{\cos({u \over a})}\nonumber\\
y &=& Y/{\cos({u \over a})}\nonumber\\
\end{eqnarray}

\noindent and setting $\Delta U \equiv {\pi a / 2},$ 
the wave metric (\ref{inwave}) becomes

\begin{eqnarray}\label{brinc}
ds^2 &=& - dU\,dV - \left({\pi \over 2\Delta U}\right)^2
(X^2 + Y^2)\,dU^2 + dX^2 + dY^2, \quad 
0 \le U \le \Delta U\nonumber\\
ds^2 &=& - dU\,dV + dX^2 + dY^2, \quad 
U < 0, U > \Delta U.
\end{eqnarray} 

This incoming wave extension of a Bertotti-Robinson spacetime is a pulse
of constant curvature of duration $\Delta U = \pi a/2$ and 
magnitude $1/a^2 = (\pi / 2\Delta U)^2.$ The focal length of the wave is 
$f = \Delta U = \pi a/2,$ meaning that null geodesics 
from an event at $U = -\infty$ 
focus at the  edge of the wave itself. The relation for colliding waves
$a b = 4 (M^2 - |\Upsilon|^2)$ is a relation between the curvatures of
the incoming waves in the $r_0 \to 0$ limit and the mean focal length
of the colliding system. It is curious that
this relationship is also enforced away from $r_0 = 0.$ 

We will compare plots of test 
particle and test string null geodesics for this metric truncated to $d=3$
in section IV.

\section{Properties of General Colliding Plane Wave Spacetimes}

\subsection{Properties of Vacuum Solutions}
In the interaction region $(u>0,v>0)$ any collinearly-polarized
colliding plane wave spacetime can be written in the form
\cite{Yurt1}:

\begin{equation}\label{basic}
ds^2 ={l_1 l_2 \over \sqrt{\alpha}}\,
e^{Q(\alpha,\beta)/2}\,
(- d\alpha^2 + d\beta^2) + \alpha {\left(
e^{V(\alpha,\beta)} dx^2 + e^{-V(\alpha,\beta)}dy^2
\right)},
\end{equation}

\noindent where the coordinate transformation from $(u,v)$ 
in equation (\ref{form}) to
$(\alpha,\beta)$ is defined by

\begin{equation}\label{alphadef}
\alpha = e^{-U(u,v)},\,\beta_u = -\alpha_u,\,
\beta_v = \alpha_v.
\end{equation}

The vacuum Einstein equations reduce to:

\begin{eqnarray}
V_{\alpha \alpha} &+& {V_{\alpha}\over \alpha} - 
V_{\beta \beta} = 0\label{Veqn}\\
Q_{\alpha} &=& - \alpha ({V_\alpha}^2 + {V_\beta}^2)\\
Q_{\beta} &=& - 2 \alpha V_\alpha\,V_\beta,\label{Qeqn}
\end{eqnarray}

\noindent plus constraints for the initial data along
$(u=0,v)$ and $(u,v=0).$ Equations (\ref{Veqn}-\ref{Qeqn}) 
are solved by functions 
whose limits as $\alpha \to 0$ are singular like

\begin{equation}\label{VQepslim}
V(\alpha,\beta) \sim \epsilon(\beta)\,\ln{\alpha} +
\mu(\beta),\qquad
Q(\alpha,\beta) \sim -\epsilon^2(\beta)\,\ln{\alpha}
+ \delta(\beta),
\end{equation}

\noindent where $\epsilon(\beta)$ is determined by an
integral over the boundary between the interaction
region and the incoming waves.

The $\beta-$dependence in the metric is ignorable if
we're only looking at the structure of the singular
terms in the metric for $\alpha \to 0.$ Counting
powers of $\alpha$ and then changing coordinates
from $(\alpha,\beta)$ to $(t,z),$ the metric behaves like 
the Kasner homogeneous, anisotropic cosmology:

\begin{equation}
ds^2 \sim -dt^2 + t^{2 p_1} dx^2 +  t^{2 p_2} dy^2 
+  t^{2 p_3} dz^2,
\end{equation}

\noindent where the Kasner exponents $\{p_i\}$ satisfy

\begin{equation}
p_1 = {2(1 + \epsilon) \over \epsilon^2 + 3},\,
p_2 = {2(1 - \epsilon) \over \epsilon^2 + 3},\,
p_3 = {\epsilon^2 - 1 \over \epsilon^2 + 3},\,
{\rm and}
\sum{p_i} = \sum{{p_i}^2} = 1.
\end{equation}

This metric has curvature squared

\begin{equation}\label{kasnerc}
R_{\mu\nu\rho\lambda} R^{\mu\nu\rho\lambda}
= {4 p_1 p_2 p_3 \over t^4},
\end{equation}

\noindent and so is singular as $t \to 0$ unless 
$\epsilon(\beta) = \pm 1,$ in which
limit the Kasner metric reduces to a slice of Rindler
spacetime.

\subsection{How This Applies to Axion-Dilaton Colliding Waves}

Axion-dilaton colliding plane waves don't obey the
vacuum Einstein equations. However, the metric obtained
through the transformation (\ref{trans})
fits the form of the metric (\ref{form}) and the coordinate
transformation (\ref{alphadef}) is still valid. (This
transformation determines the
existence of a foliation of the interaction region
into spacelike hypersurfaces $\alpha = {\rm constant}$
and works for $R_{\mu\nu}\ne 0$ as
long as the plane waves are collinearly polarized.)

Remarkably enough (but not so remarkable once one 
recalls that this is still essentially a two-dimensional 
problem), the functions $V(\alpha,\beta)$ and 
$Q(\alpha,\beta)$ still behave like (\ref{VQepslim})
in the limit $\alpha \to 0.$ Therefore, the Kasner
asymptotic limit also applies to axion-dilaton colliding
plane waves..

Combining (\ref{alphadef}) and (\ref{VQepslim}), we see
how to calculate $\epsilon(\beta)$ without integrating
over the initial data:

\begin{equation}\label{epslim}
\epsilon(\beta) = - \lim_{\alpha\rightarrow 0}
{V(\alpha,\beta)/U} 
= \lim_{\alpha\rightarrow 0}\,{{\ln{g_{xx}} - \ln{g_{yy}} 
\over \ln{g_{xx}} + \ln{g_{yy}}}}.
\end{equation}

The coordinate transformation (\ref{alphadef}) for
the metric under consideration can be solved exactly,
giving

\begin{equation}
\alpha(u,v) = {1 \over 2} \left(\cos{2 u\over a} + 
\cos{2 v \over b} \right),\quad
\beta(u,v) = {1 \over 2} \left(- \cos{2 u\over a} + 
\cos{2 v \over b} \right),
\end{equation}

\noindent and this is easily invertible to give
$(u(\alpha,\beta), v(\alpha,\beta)).$ Taking
the limit (\ref{epslim}) yields
$\epsilon(\beta) = 1,$ which means that these metrics are
in general nonsingular. However, the nonvanishing part of $V(\alpha,\beta)$ 
as $\alpha \to 0$ consists of the singular function 
$\sim \epsilon(\beta)\,\ln{\alpha}$ plus a function 
$\mu(\beta),$ which for this metric is

\begin{equation}
\mu(\beta) = \ln{{(1 - {\beta^2}) ({M^2} - {|\Upsilon|^2})
\over  {{\left( M \pm r_0 \right) }^2} -
{|\Upsilon|^2}}}.
\end{equation}

\noindent This term results in a curvature singularity in two limits: 
the Schwarzshild
limit where $r_0 = M$ and $\Upsilon = 0,$ and the extreme dilaton
limit where $r_0 = 0$ and $M = |\Upsilon|.$ 
This happens because the coordinate transformation (\ref{trans}) maps
either $r_+$ or $r_-$ to $\alpha = 0.$ The value of
$R_{\mu\nu\rho\lambda} R^{\mu\nu\rho\lambda}$ as
$\alpha \to 0$ is the same as 
$R_{\mu\nu\rho\lambda} R^{\mu\nu\rho\lambda}$
at either $r_\pm.$ The only axion-dilaton black holes where the
curvature at $r_\pm$ is not finite are the Schwarzschild
and singular dilaton black holes, with spacetime
singularities as $r \to r_-,$ as described above.

\section{Relation to Einstein-Maxwell-Dilaton Colliding Plane Wave Spacetimes}

Bret\'{o}n, Matos and Garc\'{i}a \cite{Bre} discovered a large class of colliding 
plane wave metrics that also obey the equations of motion
for the action

\begin{equation}
S = {1\over 16\pi}\int{d^4x \sqrt{-g} \{-R + 2(\nabla \Phi)^2 + 
e^{-2 \alpha \Phi} F_{\mu\nu} F^{\mu\nu} \}},
\end{equation}

\noindent which for $\alpha = 1$ is the same as the $N=1, \psi=0$ limit of
the action (\ref{action}). The metric takes the form

\begin{equation}
ds^2 = {{e^{k(\alpha,\beta)/2} \over f} 
\left(- d\alpha^2 + d\beta^2 \right)} +
({{\alpha^2} /f})\, dx^2 + f\,dy^2.
\end{equation}

For solutions that overlap with those discussed in this paper,
the function $f(\alpha,\beta)$ has the form

\begin{equation}
f={f_0 e^{\lambda(\alpha,\beta)} \over 
(a_1 \Sigma_1 +a_2 \Sigma_2)},
\end{equation}

\noindent the dilaton field is

\begin{equation}
\kappa^2 = e^{-2 \Phi} = \kappa_0^2\, (a_1 \Sigma_1 +a_2
\Sigma_2) e^{ \lambda},
\end{equation}

\noindent and the Maxwell potential has the form

\begin{equation}
A = A_{y} ={ {(a_3 \Sigma_1 +a_4 \Sigma_2)} \over {(a_1 \Sigma_1 +a_2
\Sigma_2)}}.
\end{equation}

The functions $\Sigma_{1,2}$ that reproduce the $ N=1, \psi=0$ limit of
axion-dilaton colliding waves are:

\begin{equation}
\Sigma_1 = e^{q_1\,\tau(\alpha,\beta)},\quad
\Sigma_2 = e^{- q_2\,\tau(\alpha,\beta)}. 
\end{equation}

For this class of solutions, the functions 
$\tau(\alpha,\beta),\lambda(\alpha,\beta),$
and $k(\alpha,\beta)$ satisfy

\begin{eqnarray}
\tau_{\alpha \alpha} &+& {\tau_{\alpha}\over \alpha} - 
\tau_{\beta \beta} = 0,\quad
\lambda_{\alpha \alpha} + {\lambda_{\alpha}\over \alpha} - 
\lambda_{\beta \beta} = 0,\\
k_{\alpha} &=& {\alpha \over 2} 
({\lambda_\alpha}^2 + {\lambda_\beta}^2 ) + q_1^2
({\tau_\alpha}^2 + {\tau_\beta}^2 )
,\quad
k_{\beta} =  - \alpha (\lambda_\alpha\,\lambda_\beta +
q_1^2 \tau_\alpha\,\tau_\beta),
\end{eqnarray}

\noindent and the constants $f_0,a_i,\kappa_0$ obey

\begin{equation}
f_0 \, a_1 a_2 = \kappa_0^2 \,(a_3 a_2 - a_1 a_4)^2.
\end{equation}

Since $\tau(\alpha,\beta)$ and $\lambda(\alpha,\beta)$
obey equation (\ref{Veqn}),
as $\alpha \to 0$
 
\begin{eqnarray}
\tau(\alpha,\beta) &\sim& \epsilon_1 (\beta) \ln{\alpha}\quad
\lambda(\alpha,\beta) \sim \epsilon_2 (\beta) \ln{\alpha}\\
k(\alpha,\beta) &\sim& - ( {q_1}^2 {{\epsilon_1}^2 (\beta)} + 
{{\epsilon_2}^2 (\beta)}) \ln{\alpha}
\end{eqnarray}

There are two conditions under which the metric and fields given 
above exhibit the same nonsingular Kasner
asymptotic limit as exhibited by the $N=1, \psi=0$ limit of
axion-dilaton colliding waves. The Schwarzschild limit with constant 
dilaton and Maxwell potential requires $q_1 = q_2$ and 
$\lambda(\alpha,\beta) = - q_1 \tau(\alpha,\beta),$ with 
$\epsilon_2(\alpha,\beta) = 1$ or $0.$
The Einstein-Maxwell-dilaton limit of axion-dilaton colliding waves is
reachable only if $q_2 = - q_1$ and 
$\epsilon_2(\alpha,\beta) = |q_1  \epsilon_1(\alpha,\beta)| = 1.$

\section{Maximal analytic extensions of axion-dilaton colliding 
plane waves}
In section III we showed that the asymptotic causal structure of the 
axion-dilaton colliding plane wave spacetime near the Killing-Cauchy horizon
at $u/a + v/b = \pi/2$ is that of the Kasner metric 
\begin{equation}\label{kasner}
ds^2 = -dt^2 + t^{2 p_1} dx^2 + t^{2 p_2} dy^z + t^{2 p_3} dz^2
\end{equation}

\noindent in the limit $p_1 = 1,\,p_2 = p_3 = 0,$ corresponding to
the wedges of Minkowski spacetime in Rindler coordinates that are
``behind the horizon" for the usual constantly accelerating observer. 
This insight was derived using the general asymptotic structure
of colliding plane graviational waves in \cite{Yurt1}, but it is more
easily derived using black hole coordinates. The proper time
from $r = r_\pm$ as measured by a nearby freely falling observer is
approximately

\begin{equation}
\tau_+^2(r) \sim 2 (r_+ - r) \left({r_+^2 - |\Upsilon|^2\over r_0}\right),
\quad \tau_-^2(r) \sim 2 (r - r_-) 
\left(r_-^2 - |\Upsilon|^2\over r_0\right).
\end{equation}

\noindent Changing coordinates by assigning 
$\chi_\pm =  t \, r_0/ (r_\pm^2 - |\Upsilon|^2),$ the metric becomes

\begin{equation}
ds^2 \sim - d\tau_\pm^2 + \,\tau_\pm^2 d\chi_\pm^2 + 
R(r_\pm)^2 d\Omega.
\end{equation}

\noindent In the $(\tau,\chi)$ plane the metric is the wedge of 
Rindler spacetime defined in Minkowski coordinates by
 
\begin{equation}
T^2 - X^2 = \tau_\pm^2,\quad {X\over T} = \tanh{\chi_\pm}.
\end{equation}

The axion-dilaton colliding 
plane wave maps to the wedges of
Rindler spacetime in the ``trapped regions" II and IV and the maximal
analytic extension across $\tau_\pm = 0$ gives back the parts
of Rindler space that correspond to the non-trapped regions I and III. 
It is important to remember that $\chi$ is proportional to 
$x,$ and that the
spacelike Killing vector ${\partial \over \partial x}$ 
becomes null on the Killing-Cauchy horizon at $\tau_\pm = 0.$
This signals the breakdown of spatial translation invariance in 
the $x$-direction just as ${\partial \over \partial t}$ becoming null in 
regions I and III of Fig.3b below signals the breakdown of 
time-translation invariance there. 

\vskip 0.25 cm
\centerline{\epsfxsize=6.0in\epsfbox{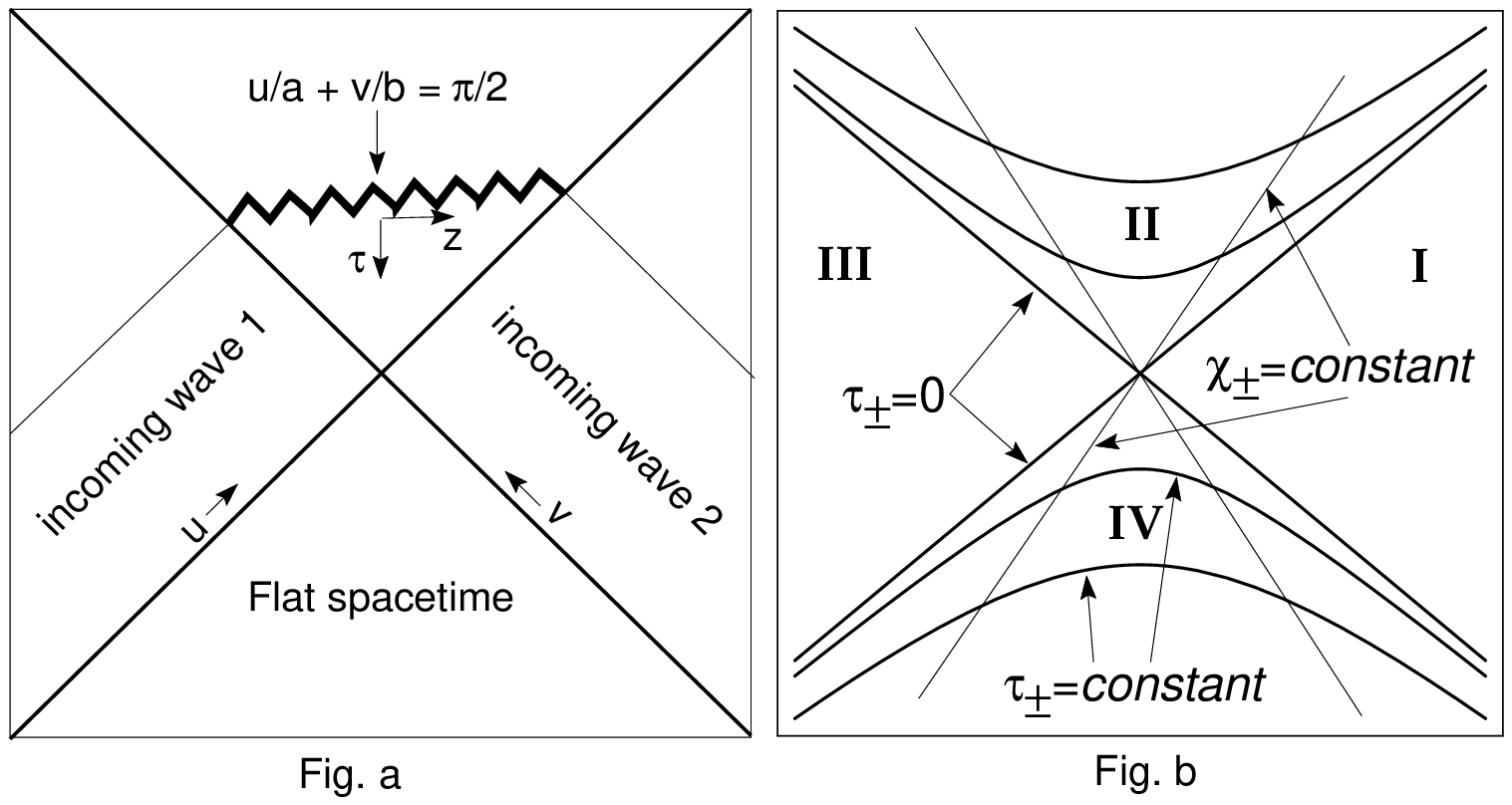}}
\noindent {\footnotesize {\sl Figs.3}
Fig.3a shows the wave collision in the $(u,v)$ or$(\tau,z)$ plane. 
Fig.3b shows how the metric
near $u/a + v/b \to \pi/2$ looks in the $(\tau,\chi)$ plane. 
The lines $\chi_\pm = const.$ are lines of
constant $x$  that cross on the Killing-Cauchy horizons $\tau_\pm = 0,$
where ${\partial \over \partial x}$ becomes null.}
\vskip 0.25 cm

From this point the maximal analytic extension of the axion-dilaton
colliding plane wave metric follows the same steps as for the 
generic axion-dilaton black hole, which has the same causal
structure and maximal analytic extension of a Reissner-Nordstrom
black hole, except in the extreme dilaton limit to which we
will return later. The Schwarzschild limit was described
by Yurtsever in \cite{Yurt1}.

There is however one problem -- we've broken the cyclic boundary 
conditions on $\phi$ in the coordinate transformation 
$\phi \to 1 +  y/{({{M^2}- {|\Upsilon|^2}} )^{1 \over 2}}.$ 
The cyclic boundary conditions on $\phi,$ as extended across the
surfaces $\tau_\pm = 0,$ can be restored by compactifying spacetime
in the $y$-direction for the incoming waves on a circle of
radius $\sqrt{{M^2}- {|\Upsilon|^2}}.$
If we insist that the maximal extension of the 
axion-dilaton colliding plane wave spacetime be analytic,
compactification of the $y$-direction is forced on the incoming waves.
\cite{Yurt1}
 
\vskip 0.25 cm
\centerline{\epsfxsize=3.0in\epsfbox{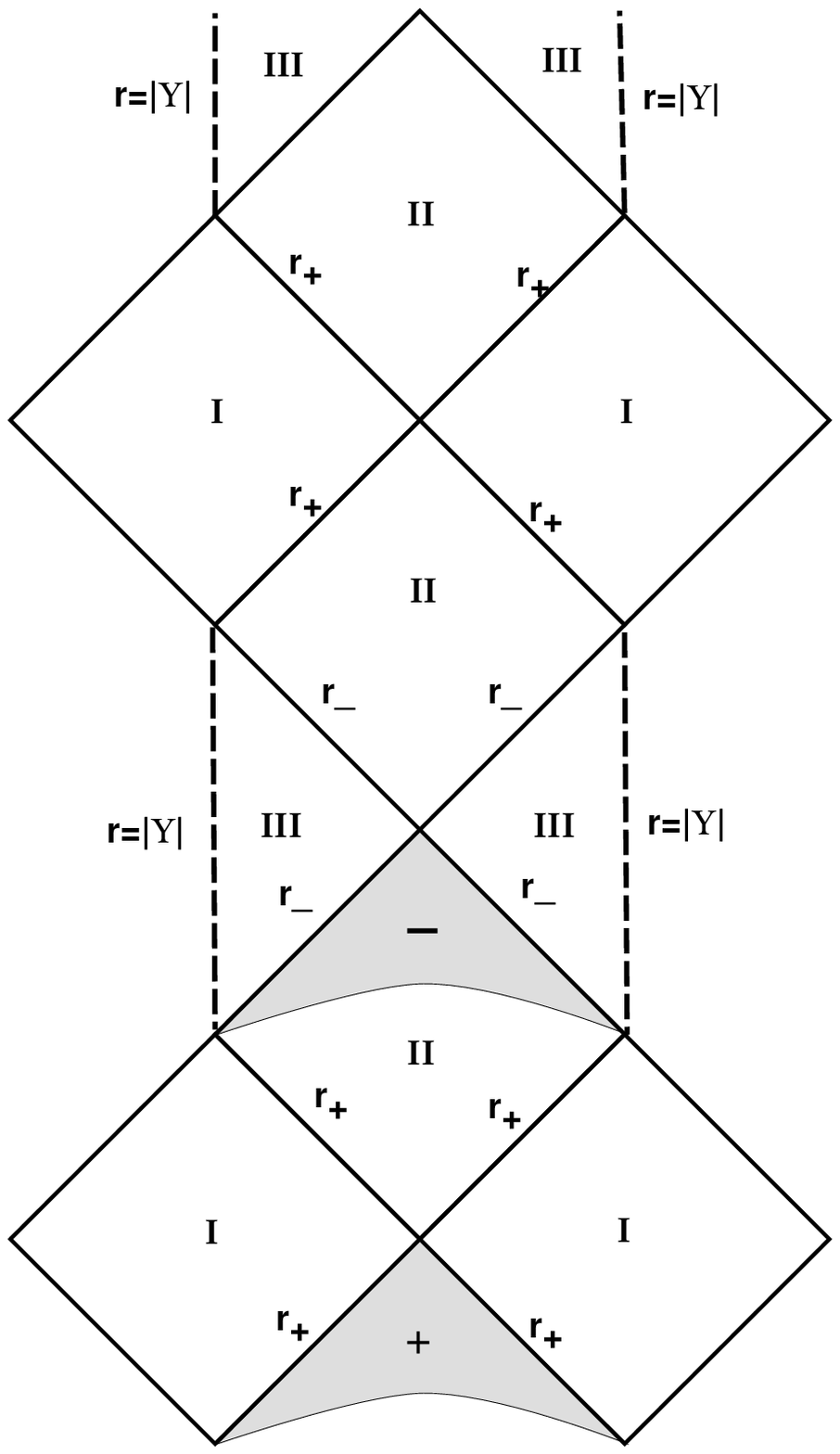}}
\noindent {\footnotesize {\sl Fig.4} The axion-dilaton
colliding plane wave metrics analytically extend
from the shaded regions of the above 
Penrose diagram into the black hole
metric above it.}
\vskip 0.25 cm

The maximal analytic extension of the 
axion-dilaton colliding plane wave metric has two sandwich waves
with translation symmtery in the $x$ and $y$ directions propagating
in a universe where the $y$-coordinate lives on a circle
of radius $\sqrt{{M^2}- {|\Upsilon|^2}}.$ The waves collide to form either
an event horizon at $r_+$ or a Cauchy horizon at $r_-$
of an axion-dilaton black hole spacetime, from which the
spacetime extends into the relevant non-trapped region of the relevant
black hole spacetime.
In the diagram above, the $(+)$ region is where the $(+)$-branch
of the axion-dilaton colliding plane wave metric extends to the
black hole spacetime to give an asymptotically flat universe plus
an axion-dilaton black hole to the future. The $(-)$ region is where
the $(-)$ branch of the colliding wave metric extends from the trapped
region II into the axion-dilaton black hole spacetime to the future.

\section{Particle vs. String propagation in this spacetime}

\subsection{How Do Test Particles Propagate Through the Focal Plane?}

The source of nearly all singularities
and causal pathologies that occur in classical general relativity
is the inevitability of the gravitational field to cause
light cones to focus in on themselves \cite{HolyBible}. 
Hawking - Penrose-type
singularity theorems chiefly express the conflict in general
relativity when the local existence and uniqueness of extremal
length curves breaks down due to the above focusing and threaten
the existence of some desired global causal structure in that
spacetime.

A simple illustration of this breakdown is gravitational lensing
with multiple images. Suppose we are looking at a spacetime where
this occurs. A light flashes at spacetime
event $E_i = (t_i,\vec{x}_i)$ and the light leaving $E_i$ is
lensed by the spacetime geometry so that an observer $O$ at
spatial location $\vec{x}_O$ sees two images of the flash
from $E_i.$ The two images seen by $O$ represent two different null
geodesics $\gamma_1$ and $\gamma_2,$ both of which leave $\vec{x}_i$
at $t=t_i.$ The geodesic $\gamma_1$ crosses $\vec{x}_O$ at $t=t_1$
and the geodesic $\gamma_2$ crosses $\vec{x}_O$ at $t=t_2.$

If $t_2 > t_1$ there is a problem. The events 
$(t_1,\vec{x}_O)$ and $(t_2,\vec{x}_O)$ cannot {\it both} lie on
the future light cone of the event $E_i,$ because the timelike observer
$O$ experiences both events. Therefore the geodesic $\gamma_1$
must lie on the future light cone of $E_i,$ while $\gamma_2$ started
out on the future light cone of $E_i$ and somehow left it.
Since the problem goes away only when $t_1 = t_2,$ it must be
true that this is where the problem starts and where null geodesics
begin to fail to determine causal boundaries in spacetime.

In general, if two null geodesics $\gamma_1$ and $\gamma_2$ intersect
once at some spacetime event $E_i$ and then reintersect at later
spacetime event $E_c,$ then both $\gamma_1$ and $\gamma_2$ leave
the ``boundary of the causal future" of $E_i$ when they cross
again at $E_c,$ and any event $E_f$ at $t_f > t_c$ along 
$\gamma_1$ or $\gamma_2$ can be reached by a timelike curve from $E_i.$

This geodesic focusing is not a problem
as long as there exists a discrete number of multiple images. 
Geodesic focusing at the continuum level is more dangerous and hence
more interesting. In general relativity the
expansion scalar $\theta$ determines when geodesic focusing
is going to interfere with the unique delimitation of causal boundaries.
If $n^a$ is a tangent vector to a null geodesic $\gamma,$
then $\theta$ is defined by $\theta = \nabla_a n^a.$ 
The rotation $\omega_{ab}$
and shear $\sigma_{ab}$ tensors are the antisymmetric and symmetric
parts of $\nabla_a n_b,$ respectively.
\footnote{These quantities are defined on the two-dimensional quotient space of vectors
orthogonal to {\bf n} modulo multiples of {\bf n}.}

The evolution equation for $\theta$
with respect to the affine parameter $\tau$ along $\gamma$ is

\begin{equation}
n^c \nabla_c\, \theta = {d\theta \over d\tau}
	= - {1 \over 2} \theta^2 - \sigma_{ab} \sigma^{ab}
	+ \omega_{ab} \omega^{ab}  - R_{cd}\,\, \xi^c \xi^d.
\end{equation}

\noindent 
Spacetimes with $\omega_{ab}\ne 0$ are not foliatable 
into spacelike hypersurfaces and hence are not stably causal, so that
term is zero if we exclude such spacetimes from consideration. Since 
$\sigma_{ab} \sigma^{ab}\ge 0,$ if $R_{cd}\,\, n^c n^d \ge 0,$
it follows that

\begin{equation}
{d\theta \over d\tau} + {1 \over 2} \theta^2 \leq 0 
\quad \rightarrow \quad
\theta^{-1} \geq {\theta_0}^{-1} + {1 \over 2} \tau.
\end{equation}

Since  $\theta \sim {1 \over V_\bot} {d V_\bot \over d\tau},$
where $V_\bot$ is the transverse volume of a bundle of 
``nearby geodesics,"
we don't want the RHS of the above inequality to cross through 
zero. If the expansion $\theta_0 < 0$ at some proper time $\tau_0$
along some geodesic,  then
$\theta\to -\infty$ along that geodesic within a proper time  
$\tau\leq 2/\vert{\theta_0}\vert$. So $V_\bot \to 0$ in a finite
amount of proper time after the bundle begins to focus, or converge,
at $\tau_0.$ When this happens to geodesics that are initially
intersecting at some previous proper time $\tau <\tau_0,$ 
either the initial value
problem breaks down or the geodesics fail to be extendible past
the focal plane. The latter alternative defines a
spacetime singularity and is accompanied by the
blowing up of curvature invariants in that region.

In the axion-dilaton colliding wave spacetime, a null vector ${\bf n}$
tangent to a null geodesic $\gamma$ can be written

\begin{equation}
{\bf n} = \dot{u} {\partial \over \partial u} + 
\dot{v} {\partial \over \partial v} +
{p_x \over g_{x x}} {\partial \over \partial x} +
{p_x \over g_{x x}} {\partial \over \partial y},
\end{equation}

\noindent where $p_x$ and $p_y$ are constants of motion along
$\gamma.$ The condition ${\bf n}\cdot {\bf n} = 0$ yields

\begin{equation}
2\,g_{uv}\,\dot{u}\,\dot{v} = - ({p_x^2 \over g_{xx}} +
{p_y^2 \over g_{yy}}).
\end{equation}

\noindent If we look at null geodesics along which $p_x = p_y = 0,$
then we can choose $\dot{v} = 0,$ and 
${\bf n} = \dot{u} {\partial \over \partial u}.$ The geodesic
equation $\ddot{u} + \Gamma^{u}_{uu}\,\dot{u}^2 = 0$ is solved
by $\dot{u} = - g^{uv},$ and the expansion scalar

\begin{equation}\label{theta}
\theta = \nabla_a n^a = 
{- 1 \over \sqrt{g}} {\partial \over \partial u} {(\sqrt{g} g^{uv})} =
{1 \over |g_{uv}|\sqrt{g_{xx} g_{yy}}} 
{\partial \over \partial u} {(\sqrt{g_{xx} g_{yy}})} 
\to - \infty 
\end{equation}

\noindent as  ${u/a} + {v/b} \to {\pi/2},$ where $a$ and $b$ are the
focal lengths defined in section II.

This focusing of
initially parallel light rays defines the Killing Cauchy
horizon on the focal plane of the collision region. Parallel
light rays delimit causal boundaries of events to the infinite
past, so information from the infinite past of the colliding
wave spacetime is focused together on the focal plane. This
spacetime is on the edge of being singular. Instead of having 
infinite curvature at the focal plane, the curvature is finite
and coordinates can be extended across it, but there is instead
the global pathology of a Killing-Cauchy horizon. Small
plane-symmetric perturbations of the incoming waves 
lead to the generic singular  solutions.

Note that $V_\bot = \sqrt{g_{xx} g_{yy}} = 
|\cos{({u \over a} + {v \over b})} 
\cos{({u \over a} - {v \over b})}|$ is independent of $r_0,$
so the focusing is controlled by the
supersymmetric limit $r_0 \to 0$ of
the Bertotti-Robinson colliding plane wave
system. Therefore comparisons of test particle and test string
propagation can be made using the incoming wave extended from
the Bertotti-Robinson collision region via the Khan-Penrose
prescription, and the results should apply to axion-dilaton
colliding plane waves with $r_0 \ne 0.$

In order to plot these geodesics, it is convenient to
truncate the metric (\ref{bert}) to $d=3$. Changing to
harmonic coordinates gives the metric

\begin{eqnarray}\label{bert2}
ds^2 = dU \, dV + h(U) X^2\, dU^2 - dX^2, 
\qquad h(U) &=& ({\pi \over 2 \Delta U})^2 \quad 0 < U < {\Delta U},\nonumber\\
&=& 0 \quad U < 0,\,\, U > {\Delta U},
\end{eqnarray}

\noindent where $\Delta U = {\pi a /2}.$ The geodesic equations are

\begin{equation}
\ddot{V} + {\partial h\over \partial U}\,X^2\,\dot{U}^2 
+ 4 h(U)\, \dot{U}\, X \dot{X} = 0,\quad
\ddot{U} = 0, \quad \ddot{X} + h(U)\, \dot{U}^2\, X = 0,
\end{equation}

\noindent and the null condition gives 

\begin{equation}
\dot{U}\,\dot{V} + h(U) X^2\,\dot{U}^2 - \dot{X}^2 = 0.
\end{equation}

The above equations are invariant under rescaling the affine 
parameter by $\tau \to \alpha \tau,$ so the paths
of massless test particles are the same for particles of all energies,
a general feature of Einstein relativity. Therefore it is convenient and 
proper to choose for the above spacetime $U = \tau,$ after which the
equations are easily solved. Null geodesics passing through this wave
take the form 

\begin{eqnarray}
U < 0\qquad X(\tau) &=& p_0\,\tau,\quad 
V(\tau) = p_0^2\,\tau,\nonumber\\
0 < U < {\Delta U}\qquad X(\tau) &=& c_0\,\sin{(\omega_0\,\tau)}
+ d_0\,\cos{(\omega_0\,\tau)},\nonumber\\
V(\tau) &=& \int{\dot{X}(\tau)^2\,d\tau} + v_{02},\nonumber\\
U > {\Delta U}\qquad X(\tau) &=& p_f\,\tau + x_f,\quad 
V(\tau) = p_f^2\,\tau + v_f,
\end{eqnarray}

\noindent where $\omega_0 = {\pi \over 2 \Delta U}$ 
and the parameter $p_0$ represents
the test particle momentum in the $x$-direction. The six constants above
are determined by the continuity of $X(\tau),$ $\dot{X}(\tau),$ and
$V(\tau)$ (but not $\dot{V}(\tau)$) across the surfaces $U=0$ and
$U=\Delta U.$ The geodesics were plotted below using {\sl Mathematica}. 

\vskip 0.25 cm
\centerline{\epsfxsize=3.0in\epsfbox{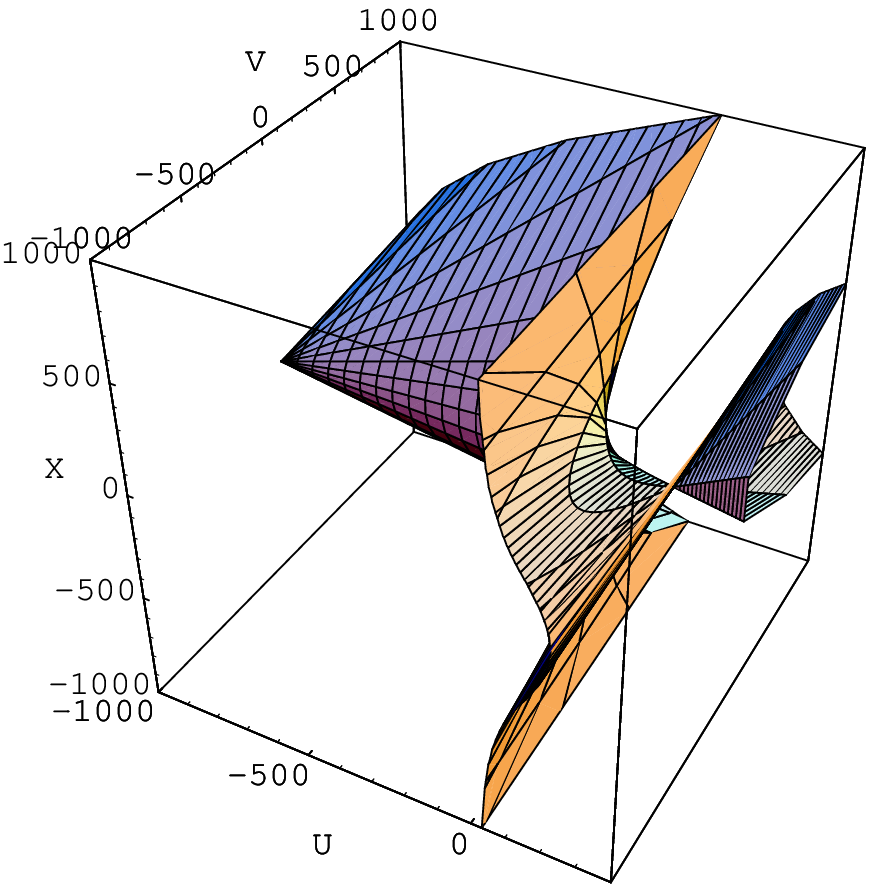}}
\noindent {\footnotesize {\sl Fig.5}\, Null geodesics from 
$(-1000,0,0)$ pass through the wave between $U=0$ and $U=200$
and are focussed to a point. A similar picture was shown
by Penrose in \cite{Penwave}.}
\vskip 0.25 cm

In figure $5$ the plane wave passes
between $U=0$ and $U=200.$
After the null geodesics focus at $U\sim 216,$ they fail to determine
the boundary of the causal future of the initial event,
and the light cone is expanded
out along the direction {\sl parallel} to the wave. Null geodesics from
an event at $U= -\infty$ would focus exactly at $f = \Delta U = 200.$

Because of the extreme distortion of the light cone by the plane 
wave, every spacelike hypersurface in this spacetime intersects 
at least one null geodesic more than once. A global Cauchy surface
cannot be defined, but for local calculations one can define a partial
Cauchy surface and compute field theory Bogolyubov
coefficients.
Gibbons \cite{Gib1} showed that although the quantum theory of a 
scalar field in a single plane wave background is
easily calculable and yields no particle creation, the theory itself
becomes singular at the focal plane where the Cauchy horizon
can no longer be neglected.

\subsection{How Do Test Strings Propagate Through the Focal Plane?}

Plane gravitational waves are interesting string backgrounds
to explore because the metric fields provide exact conformally
invariant couplings on the string world sheet. This is because
all the higher-derivative terms that could add (worldsheet) quantum corrections
vanish identically.\cite{Amkl}
String propagation through gravity waves has been fruitfully explored
in the past in the context of scattering amplitudes. A notion of
``stringy singularity" based on infinite string excitation was
examined by Horowitz and Steif \cite{Horo1,Horo2}, 
Sanchez and de Vega \cite{SDV1} and others.
While this looks like a good operational definition of singular string
propagation, it doesn't shed light on the nature of causal volume
delimitation in string theory and the potential physically-relevant
pathologies that could occur when causal volumes are delimited
by solutions to worldsheet rather than wordline mathematics.
For this reason, we step back to that earlier work and re-examine
it from a geodesic rather than an S-matrix point of view.

In extending the geodesic picture to string theory, the test particle 
geodesics that define the boundary of the test particle light cone
are represented by the zero mode of the string. This is the center-of-mass
coordinate that obeys that standard geodesic equation. If we only
look at the geodesics of test string zero modes, then the singularities
and causal pathologies of general relativity remain with minor modifications
(in the cases where we trust the background spacetime approximation,
at least.)

This is basically telling us that test particles propagate in ``stringy
general relativity" rather similarly to how they propagate in ordinary
general relativity. The biggest difference comes from the rescaling of
the stringy affine parameter relative to the test particle affine
parameter by $e^{-2\phi}.$ This has a noticeable effect mainly
in the case of a dilaton black hole with purely electric charge.
\cite{Horo1}

If we take all string modes into account, the counterpart
to a geodesic equation in string theory becomes
\begin{equation}
\square X^\mu + \Gamma^{\mu}_{\nu \lambda} 
(X(\tau, \sigma )) \partial_a X^\nu \partial^a X^\lambda = 0.
\end{equation}
 
\noindent In the single plane wave metric (\ref{bert}) 
the equations reduce to

\begin{eqnarray}
\ddot{V} - {V}'' + {\partial h\over \partial U}\,X^2\,
(\dot{U}^2 - {U'}^2)
+ 4 h(U)\,X\,(\dot{U}\dot{X} - U' X') = 0,\nonumber\\
\ddot{U} - U'' = 0, \quad 
\ddot{X} - X''+ h(U)\, (\dot{U}^2 - {U'}^2)\, X = 0.
\end{eqnarray}

The mass shell constraints come from the vanishing of the worldsheet
stress tensor and automatically satisfy the first equation above.
If we choose the gauge $U = U(\tau)$ we get 

\begin{equation}\label{constraint}
\dot{U}\,\dot{V} = - h(U) X^2\, \dot{U}^2 + (\dot{X}^2 + {X'}^2), 
\quad \dot{U} V' = 2\,\dot{X}\,X'
\end{equation}
 
\noindent and the remaining second order equation reduces to

\begin{equation}
\ddot{X} - X''+ h(U)\, \dot{U}^2\, X = 0.
\end{equation}

These equations don't allow the rescaling of string proper affine parameter,
so if we further fix the gauge by $U = p \tau$ and try to rescale
$p$ out of the equations through $\tau' = p \tau,$ factors of $p$ end 
up in the $X'$ terms. Setting $L_{string} = 1$ and expanding in open 
string modes using 
$X(\tau,\sigma) = \sum{X_n(\tau)} \cos{(n \sigma)},$ we get

\begin{equation}
\ddot{X}_n(\tau) = 
- \left({n^2 \over p^2} + h(U)\right)\, X_n(\tau),
\end{equation}

\noindent with $V(\tau,\sigma) = \sum{V_n(\tau)} \cos{(n \sigma)}$
obtainable by straightforward integration of (\ref{constraint}). As pointed out in \cite{Mende}, string
null trajectories are momentum-dependent and hence fail
to satisfy the Principle of Equivalence observed in
particle geodesics. So causal boundaries as determined by
propagating strings become momentum-dependent.

Assigning $\omega_n = \sqrt{{n^2 \over p2 } + h_0}$
and $\omega_0 = \sqrt{h_0} = \pi/{2 \Delta U},$ it is convenient to expand in 
the basis:

\begin{eqnarray}
U < 0\quad X_0(\tau) &=& p_0\,\tau,\quad 
X_n(\tau) = a_n \cos{(n \tau/p)} + b_n \sin{(n \tau/p)}\\
0 < U < {\Delta U}\quad X_0(\tau) &=& c_0\,\sin{(\omega_0\,\tau)}
+ d_0\,\cos{(\omega_0\,\tau)},\\
X_n(\tau) &=& c_n\,\sin{(\omega_n\,\tau)}
+ d_n\,\cos{(\omega_n\,\tau)}\\
U > {\Delta U}\quad X_0(\tau) &=& p_f\,\tau + x_f,\quad 
X_n(\tau) = e_n \cos{(n \tau/p)} + f_n \sin{(n \tau/p)}
\end{eqnarray}

\noindent with $U = \tau$ and the $V_n(\tau)$ obtained by integrating (\ref{constraint}).

\noindent This is related to the more common expansion for 
strings in flat spacetime 

\begin{equation}
U = p \tau,\quad X(\tau,\sigma) = X_0(\tau) + i\sum_n{{\alpha_n\over n}\,
e^{- i n \tau}\, cos(n \sigma)}
\end{equation}
through 

\begin{equation}\label{basis}
a_n = -{2 p\over n} {\rm Im}\,\alpha_n,\quad 
b_n = {2 p\over n} {\rm Re}\,\alpha_n
\end{equation}
 
Applying continuity equations across the waves boundaries at
$U=0$ and $U=\Delta U$ gives the linear transformation between
incoming and outgoing mode constants $(a_n,b_n)$ and $(e_n,f_n)$:

\begin{eqnarray}
e_n = &a_n&\,\{\cos(n \Delta U/p) \cos(\omega_n\,\Delta U) + 
({\omega_n\,p\over n}) \sin(n \Delta U/p) \sin(\omega_n\,\Delta U)\}\nonumber\\
+ &b_n&\,\{-\sin(n \Delta U/p) \cos(\omega_n\,\Delta U) + 
({n \over \omega_n\,p}) \cos(n \Delta U/p) \sin(\omega_n\,\Delta U)\},\\
f_n = &a_n&\,\{\sin(n \Delta U/p) \cos(\omega_n\,\Delta U) - 
({\omega_n\,p\over n}) \cos(n \Delta U/p) \sin(\omega_n\,\Delta U)\}\nonumber\\
+ &b_n&\,\{\cos(n \Delta U/p) \cos(\omega_n\,\Delta U) + 
({n \over \omega_n\,p}) \sin(n \Delta U/p) \sin(\omega_n\,\Delta U)\}.
\end{eqnarray}

\noindent Transforming back to the basis $(\alpha_n,\alpha_{-n})$
by undoing (\ref{basis}), the Bogolyubov coefficients 
$B_n$ obtained match those obtained for $d=4$ in 
\cite{JN}, which according to the conventions used here is

\begin{equation}\label{Bog}
|B_n|^2 = {1 \over 4} \left({p \over n \omega_n}\right)^2 
\omega_0^4 \,
\sin^2(\omega_n \Delta U).
\end{equation}

\noindent It is significant that this coefficient is zero in
scalar quantum field theory \cite{Gib1}. As Gibbons explained,
there is no mixing between in and out bases in that case because there 
is a global null Killing vector guaranteeing that frequencies
can be measured in the same way before and after the wave's passage.
Strings are excited because they have extended structure. String in and out bases are getting mixed
in outright defiance of this target space Killing vector that has
such a powerful restrictive effect on quantum fields. 

The limit $\Delta U \to 0$ leads to a wave profile
$h(U) \to {\pi \over 2} \delta(U),$ which in \cite{Horo1} and
\cite{JN} was shown for bosonic strings 
to satisfy the definition of a singularity
in terms of string propagation because the mass operator for the
``out" state in the ``in" vacuum diverges like $\sum{1 \over n}$.

For the single wave under consideration  $\Delta U = \pi a/2.$
The axion-dilaton colliding wave metric requires 
$a b = 4(M^2 - |\Upsilon|^2),$ so the limit in which
one or both incoming waves has the profile 
$h(U) \to {\pi \over 2} \delta(U)$ is also the limit in which
the maximal analytic extension of the collision region
gives an extreme dilaton black hole with zero entropy but
infinite curvature at the horizon and $1/2$ of
the $N=4$ supersymmetry unbroken. \cite{RK1}

String motion through the wave represented by (\ref{bert}) looks
the same globally as the particle motion  when plotted at
the same scale as in figure $5.$ The main difference becomes visible in the focusing
region when the momentum is varied, as shown below:

\vskip 0.25 cm
\centerline{\epsfxsize=6.5in\epsfbox{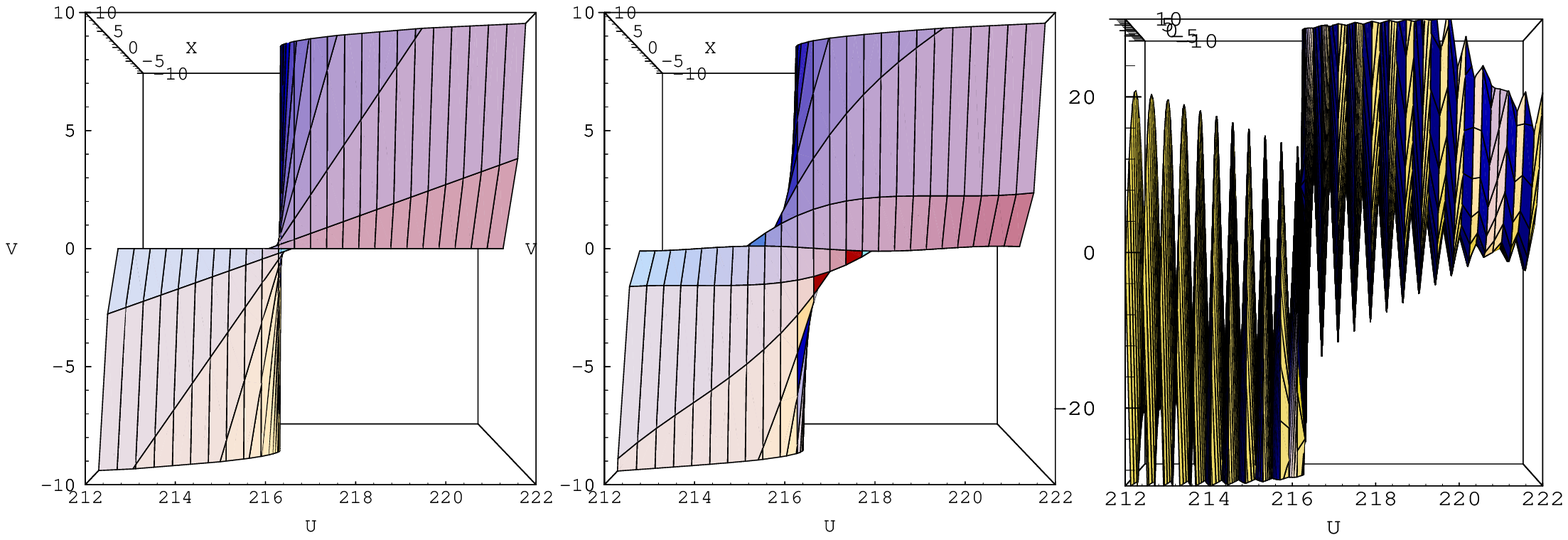}}
\noindent {\footnotesize {\sl Fig.6}\, The surface
swept out by $X(\tau,0)$ plotted for: $p = 1000, p=2$ and $p = .01.$}
\vskip 0.25 cm

The plot above shows that the focal region as determined by
strings becomes smeared by strings as
the momentum decreases. This does not mean that 
string trajectories are
no longer leaving the boundary of the causal future after they cross.
This still has to be true at large distances. String effects obscure
the location of the focal plane but not the effects of geodesic
focusing itself.

The geodesic focusing that determines the location of the focal plane
of the extreme single plane wave in (\ref{bert}) was shown in 
equation (\ref{theta})
to control the focusing of the Killing-Cauchy horizon
in the collision region as $u/a + v/b \to \pi/2.$ The Killing-Cauchy
horizon for axion-dilaton colliding plane wave system is mapped to
$r= r_\pm$ in the axion-dilaton black hole via the
coordinate transformation (\ref{ctrans}). The quantity
$V_\bot = \sqrt{g_{xx}\,g_{yy}}$ is locally identified with
$\sqrt{-g_{tt}\,g_{\phi\phi}}.$ This demonstrates a relationship
between infinite geodesic focusing in the colliding wave system and
the infinite red shift in the black hole system. 

Consider the ``stringy stretched horizon" as elucidated in \cite{Suss1}.
An observer accelerating at constant $r$ (a FIDO)
near a black hole event horizon 
sees a passing freely falling string with a time resolution that decreases like
$\varepsilon \sim e^{- c t},$ with 
$c = (r_+ - r_-)/2 (r_+^2 - |\Upsilon|^2).$ But we know that 
a measurement of the size of a string cut off at mode N grows like 
$\log{N},$ $i.e.$ strings fill more space as we try to measure
them with greater time resolution.
Since the resolution $\varepsilon \sim 1/N$, the FIDO
would see the passing string begin to grow like $c t$ until
it filled the horizon area. In \cite{Suss1} the authors
fixed $p = 1$ and looked at $\varepsilon(N).$ 
In the plots in figure $6$ we fixed $N = 1$ and varied
$p$ instead, finding that as we try to look
at the string with decreasing resolution $\varepsilon = p/N,$
the string gets longer and fills more space.

The Killing-Cauchy horizon formed by axion-dilaton colliding
plane waves maps to a {\it past} horizon of an axion-dilaton
black hole. (See figure $4.$) 
So the ``stringy stretched focal plane" can be viewed as the 
time-reversed version of the ``stringy stretched horizon" 
described in \cite{Suss1}. In other
words, suppose we are in the maximally extended colliding plane
wave spacetime described in section V,
where two waves in a cylindrical universe collide to produce
the axion-dilaton black hole spacetime in figure $4$ at $r = r_+.$
A FIDO close to $r_+$ in region I would see a test string emerging
from the collision region at $t = -\infty$ filling the past 
horizon of the white hole created by the collision
and then shrinking rapidly. This is the time-reversed 
version of what the FIDO at the future horizon sees.

It is important to remember, however, that these nonsingular
colliding plane wave metrics are unstable. The
singular term in the curvature (\ref{kasnerc}) only vanishes when
the product $p_1 p_2 p_3$ is precisely zero everywhere, which
only happens if the initial data is specified with an arbitrarily
high precision. 

\subsection{Scattering of Almost-Plane Waves}

It was shown by Yurtsever \cite{Yurt3,Yurt2} that two
finite-sized gravity waves that are ``nearly
plane symmetric" over some transverse size 
$L_T \sim L_{1 T} \sim L_{2 T}$ will
collapse through plane-symmetric processes if the average
focal length $f = \sqrt{f_1 f_2}$ of the incoming waves
satisfies $L_T \gg f.$ There is a causality-based
argument for this: The proper time in the collision region
for the singularity or Cauchy horizon to form is 
$\Delta \tau \sim f.$ So if $L_T \gg f,$ the asymptotic evolution
becomes dominated by $infinite$ plane wave dynamics before
gravitational shock waves containing the information that the 
incoming waves are $finite$
in extent could have time to reach the collision region.

The mass-energy density contained in each incoming wave of thickness
$a_i$ and average curvature $\sim R_i$ would be on the order of
$E_i/(a_i L_T^2) \sim R_i.$ The focal length $f_i \sim a_i/(a_i^2 R_i),$
which gives $E_i \sim L_T^2/f_i.$ (So the mass-energy per unit
area in a finite, nearly-plane symmetric gravity wave is
$E_i/A_i \sim 1/f_i.$)
The total mass-energy in the collision
region then would be 
$E_{CW} \sim \sqrt{E_1 E_2} \sim L_T^2/\sqrt{f_1 f_2} = L_T^2/f,$
so the condition $L_T \gg f$ implies  $E_{CW}\gg L_T.$ 
In other words, the mass-energy in the colliding wave system is
contained well within its Schwarzschild radius when the two waves 
meet and the final product of this collision ought to be a black 
hole of size $\sim L_T^2/f.$

In the case of axion-dilaton colliding waves, these spacetimes are
in general nonsingular and hence are believed to be unstable, in
that small plane-symmetric perturbations on the initial data
propagate to cause the Killing-Cauchy horizon to become singular.
However, $f =\pi \sqrt{a b}/2 = \pi \sqrt{M^2 - |\Upsilon|^2}.$
For the incoming waves, $R_i \sim 1/a_i^2.$ Therefore
$f_i \sim a_i,$ and the limit $f_i \to 0$ is also the limit
$R_i \to \infty.$
The condition $L_T \gg f$ implies  
$E_{CW}\gg L_T\gg f=\pi \sqrt{M^2 - |\Upsilon|^2}.$ This
suggests that the collision of these finite waves could nucleate
not one, but several axion-dilaton black holes, and in the maximally supersymmetric limit of $M \to |\Upsilon|,$ the result could be
an explosion of extreme dilaton black holes, which aren't
really black holes because the event horizon is singular. 
For that to happen at
least one of the incoming waves would have zero thickness and
infinite curvature. Such an incoming wave is already singular if
we use the operative definition of a singular wave in string
theory as a background in which the Bogolyubov coefficient for string
excitation becomes infinite.

\section{Conclusions}

The local coordinate transformation between the trapped region of a Schwarzschild black hole and a colliding plane gravitational wave discovered by Ferrari and Iba\~{n}ez \cite{FerIban1,FerIban2} extends naturally to the
class of axion-dilaton black holes that are classical solutions to the
electric-magnetic duality-invariant action (\ref{action})

\begin{eqnarray}
S_{eff} = {1\over 16\pi}\,\int d^4 x
\, \sqrt{-g}\,\, (
- R  + {1\over 2} {\partial_\mu \lambda \partial^\mu \bar{\lambda} \over 
({\rm Im}\lambda)^2} 
- \sum_{n=1}^{N}{F^{(n)}_{\mu\nu} {}^\star \tilde{F}^{(n)}{}^{\mu\nu}}).
\nonumber
\end{eqnarray}

\noindent The local coordinate transformation (\ref{ctrans})

\begin{eqnarray}
r &\to& M \pm r_0\, \sin ({u\over a} + {v\over b}),\qquad
\theta \to {\pi \over 2} \pm \left(
{u \over a} - {v \over b}\right),\nonumber\\
t &\to& x\,{r_0}/{({{M^2}- {|\Upsilon|^2}} )^{1 \over 2}},
\qquad
\phi \to 1 +  y/{({{M^2}- {|\Upsilon|^2}} )^{1 \over 2}}
\nonumber
\end{eqnarray}

\noindent transforms an axion-dilaton black hole metric characterized by 
mass $M$ and complex axion-dilaton charge $\Upsilon$ to the collision
region of a colliding axion-dilation plane wave metric (\ref{adcwmetric})

\begin{eqnarray}
g_{uv} &=& {{-2\,\left({{{\left( M \pm 
          r_0\,\sin ({u\over a} + {v\over b})
           \right) }^2} -{|\Upsilon|^2}} \right)}
\over a b},\nonumber\\
g_{xx} &=& {{\left({{ M^2} - {|\Upsilon|^2}}\right)\,
{{\cos ({u\over a} + {v\over b})}^
       2}}\over 
   {{{\left( M \pm 
           r_0\,\sin ({u\over a} + {v\over b})
           \right) }^2} -{|\Upsilon|^2} }}\nonumber\\
g_{yy} &=& {\cos^2 ({u\over a} - {v\over b})}\,
     {\left( {{{\left( M \pm 
          r_0\,\sin ({u\over a} + {v\over b})
           \right) }^2} -{|\Upsilon|^2} } \right) \over 
   {{ M^2} - {|\Upsilon|^2}}}.\nonumber
\end{eqnarray}

\noindent The constants $a$ and $b$ represent the focal lengths of the
incoming waves obtained from above through the Khan-Penrose prescription
\cite{KhanPen} and satisfy the relation 
$a b = 4 (M^2 - |\Upsilon|^2).$ This metric has a Killing-Cauchy horizon
at ${u \over a} + {v \over v} = {\pi \over 2},$ 
where the spatial translation Killing
vector ${\partial \over{\partial x}}$ becomes null. The curvature at
the Killing-Cauchy horizon
is equal to the curvature at $r = r_\pm$ of the
correspoonding axion-dilaton black hole and so is finite except in the
Schwarzschild and extreme electrically or magnetically charged
dilaton limits where the curvature at $r_-$ diverges.

The limit $r_0 \to 0,$ which for the black hole metrics corresponds
to an extreme black hole, takes the axion-dilaton colliding plane wave
metric to the Bertotti-Robinson metric (\ref{bert})

\begin{eqnarray}
ds^2 = - du\,dv + {{\cos^2({u\over a} + {v\over b})}}\,
dx^2 + {\cos^2 ({u\over a} - {v\over b})}\,dy^2,\nonumber
\end{eqnarray}

\noindent which has a finite average focal length 
$a b = 4 (M^2 - |\Upsilon|^2)$ despite fact that the trapped
region of the corresponding black hole has become infinitesimal. The product $a b$ of the non-vanishing
parameters describing the colliding waves is related
to the entropy of a nonsingular extreme black hole
with $1/4$ unbroken $N=4$ supersymmetry through

\begin{equation}
a b = {4 S_{extr}\over \pi}.
\end{equation}

An incoming wave obtained from the Bertotti-Robinson collision
region can be described in harmonic coordinates as a shock
wave of thickness $\Delta U = \pi a/2,$ where $a$ is the
focal length of that wave, and constant curvature of 
magnitude $1/a^2 = (\pi / 2\Delta U)^2.$ If we send $a \to 0$
while keeping the other incoming focal length $b$ finite, then
the constraint $a b = 4 (M^2 - |\Upsilon|^2)$ says that
$M = |\Upsilon|.$ The limit $a \to 0$ corresponds to a delta
function incoming wave. The black hole corresponding to the
$M = |\Upsilon|$ limit has a singular horizon, zero entropy and
$1/2$ of $N=4$ supersymmetry unbroken. 
This correspondence between an
delta-function gravity wave and this extreme dilaton
configuration with zero entropy 
is like a type of wave-particle duality in string theory,
albeit not the usual one.

The maximal analytic extension of the metric (\ref{adcwmetric})
across the Killing-Cauchy horizon gives back the non-trapped regions
of the corresponding axion-dilaton hole, but requires that
the $y$ coordinate live on a circle of radius 
$\sqrt{M^2 - |\Upsilon|^2}.$ The resulting spacetime has two 
plane-symmetric single waves propagating in a cylindrical universe
that collide and form a past horizon of an axion-dilaton black
hole, shown in figure 4.

The propagation of test particle and test strings in a plane
gravitational wave were compared. Geodesic focusing for the
axion-dilaton colliding wave system is controlled by the
supersymmetric Bertotti-Robinson limit. The single plane
waves obtained from this collision region metric therefore
make good toy backgrounds to study stringy geodesic focusing.
The string equivalent of a massless geodesic equation does not
allow for rescaling the affine parameter; consequently
light cones as delimited by strings depend on momentum. 
This introduces a time resolution dependence into string geodesic focusing
that is the same time resolution dependence that was 
analyzed in the stretched black hole horizon by Susskind in 
\cite{Suss1}, suggesting that a ``stretched focal plane" is the
colliding plane wave analog of a stretched horizon for the black hole.

\section{Acknowledgments}

The author gratefully acknowledges discussions with Renata Kallosh,
John Schwarz, Gary Gibbons and Gary Horowitz 
during the preparation of this paper.

\end{document}